\documentclass[prd,twocolumn,amsfonts,amssymb]{revtex4}

\usepackage{graphicx}

\usepackage{bm}

\newcommand{\beq}{\begin{equation}}
\newcommand{\eeq}{\end{equation}}
\newcommand{\beqs}{\begin{eqnarray}}
\newcommand{\eeqs}{\end{eqnarray}}

\begin{document}

\title{Infrared Zero of $\beta$ and Value of $\gamma_m$ for an SU(3) 
Gauge Theory at the Five-Loop Level} 

\author{Thomas A. Ryttov$^a$ and Robert Shrock$^b$}

\affiliation{(a) \ CP$^3$-Origins and Danish Institute for Advanced Study \\
Southern Denmark University, Campusvej 55, Odense, Denmark}

\affiliation{(b) \ C. N. Yang Institute for Theoretical Physics \\
Stony Brook University, Stony Brook, NY 11794, USA }

\begin{abstract}

We calculate the value of the coupling at the infrared zero of the beta 
function of an asymptotically free SU(3) gauge theory at the five-loop level 
as a function of the number of fermions.  Both a direct analysis of the beta
function and analyses of Pad\'e approximants are used for this purpose. We then
calculate the value of the five-loop anomalous dimension, $\gamma_m$, 
of the fermion bilinear at this IR zero of the beta function.

\end{abstract}

\pacs{11.15.-q,11.10.Hi,11.15.Bt}

\maketitle

The evolution of an asymptotically free gauge theory from the ultraviolet (UV)
to the infrared (IR) is of fundamental importance. The evolution of the running
gauge coupling $g=g(\mu)$, as a function of the Euclidean momentum scale,
$\mu$, is described by the renormalization-group (RG) beta function \cite{rg},
$\beta_g = dg/dt$ or equivalently, $\beta = d\alpha/dt = [g/(2\pi)] \,
\beta_g$, where $\alpha(\mu) = g(\mu)^2/(4\pi)$ and $dt=d\ln \mu$ (the argument
$\mu$ will often be suppressed in the notation).  Here we consider a vectorial
gauge theory with gauge group $G={\rm SU}(3)$ and $N_f$ flavors of fermions
$\psi_i$, $i=1,...,N_f$ transforming in the fundamental (triplet)
representation.  We impose the condition of asymptotic freedom (AF) for the
self-consistency of the perturbative calculation of $\beta$.  For simplicity,
we take the fermions to be massless \cite{fm}.  This theory is quantum
chromodynamics (QCD) with $N_f$ massless quarks.

The beta function of this theory has the series expansion
\beq
\beta = -2\alpha \sum_{\ell=1}^\infty b_\ell \, a^\ell =
-2\alpha \sum_{\ell=1}^\infty \bar b_\ell \, \alpha^\ell \ ,
\label{beta}
\eeq
where $a=g^2/(16\pi^2)=\alpha/(4\pi)$, $b_\ell$ is the $\ell$-loop coefficient,
$\bar b_\ell = b_\ell/(4\pi)^\ell$, and we extract a minus sign for
convenience. The $n$-loop ($n\ell$) beta function, denoted $\beta_{n\ell}$, is
obtained from Eq. (\ref{beta}) by changing the upper limit on the $\ell$-loop
summation from $\infty$ to $n$.  The (scheme-independent) one-loop and two-loop
coefficients are $b_1=11-(2/3)N_f$ \cite{b1} and 
$b_2=102-(38/3)N_f$ \cite{b2}. The AF condition
implies the upper bound $N_f < N_{f,b1z}=33/2$ \cite{nfintegral}, i.e., 
the integer upper bound $N_f \le 16$, which we impose. 
We denote the interval $0 \le N_f \le 16$ as $I_{AF}$. The $b_\ell$ with 
$\ell \ge 3$ are scheme-dependent \cite{gross75}; 
$b_3$ and $b_4$ were calculated in \cite{b3}
and \cite{b4} (and checked in \cite{b4p}), 
in the $\overline{\rm MS}$ scheme \cite{msbar}; e.g., $b_3 =
(2857/2)-(5033/18)N_f+(325/54)N_f^2$. 
As $N_f \in I_{AF}$ increases from 0, $b_2$ decreases, vanishing at 
$N_{f,b2z}=153/19=8.05$, and is negative in the real interval 
$153/19 < N_f < 33/2$, i.e., the integer interval 
$I_{IRZ}: \ 9 \le N_f \le 16$. If $N_f \in I_{IRZ}$, then the two-loop
beta function $\beta_{2\ell}$ has an IR zero (IRZ), at
$\alpha=\alpha_{IR,2\ell}=-4\pi b_1/b_2$.  Here we denote the IR zero (if it
exists) of the $n$-loop beta function $\beta_{n\ell}$ as $\alpha_{IR,n\ell}$.
For $N_f$ near the upper end of $I_{IRZ}$, $\alpha_{IR,2\ell}$ is small and can
be studied perturbatively \cite{b2,bz}. As $N_f \in I_{IRZ}$ decreases, 
$\alpha_{IR,2\ell}$ increases toward strong coupling.  Hence, to
study the IR zero for $N_f$ toward the middle and lower part of $I_{IRZ}$ 
with reasonable accuracy, one requires higher-loop calculations. These
were carried out to four-loop order in \cite{gk}-\cite{bc} \cite{irgen}.
Clearly, such a perturbative 
calculation of the IR zero of $\beta_{n\ell}$ is only reliable if the resultant
$\alpha_{IR,n\ell}$ is not excessively large. 
Since the $b_\ell$ with $\ell \ge 3$ are scheme-dependent, it is
necessary to assess the sensitivity of the value obtained for
$\alpha_{IR,n\ell}$ for $n \ge 3$ to the scheme used for the calculation. This
was done in \cite{sch}-\cite{gracey2015} 
(see also \cite{tr2016,stevenson2016}). In
\cite{sch}-\cite{sch2}, a set of conditions that an acceptable scheme
transformation must satisfy were presented, and it was shown that although
these are automatically satisfied in the local vicinity of the origin,
$\alpha=0$ (as in optimized schemes for perturbative QCD calculations
\cite{qcdschemes,brodsky}), they are not automatically satisfied, and indeed,
are quite restrictive conditions, when one applies the scheme transformation at
an IR zero away from the origin.

Here we report the first calculation of the five-loop IR zero 
of $\beta$ and resultant five-loop evaluation of the 
anomalous dimension of the fermion bilinear at this IR zero, 
for $N_f \in I_{IRZ}$, making use of the recent calculation of $b_5$ 
in the $\overline{\rm MS}$ scheme from \cite{bck}.  The results are of
fundamental importance for understanding the RG evolution of SU(3) gauge theory
with variable fermion content. 

The anomalous dimension $\gamma_m$ of the fermion 
bilinear operator $\bar\psi_i\psi_i$ (no sum on $i$) is defined as 
$D(\bar\psi_i\psi_i) = 3-\gamma_m$, where $D$ is the full scaling dimension.  
Knowing $\alpha_{IR,n\ell}$, one can then evaluate $\gamma_m$
(calculated to the same $n$-loop order) at $\alpha=\alpha_{IR,n\ell}$; we
denote this as $\gamma_{IR,n\ell}$.  This anomalous dimension is of particular
interest, since (if calculated to all orders) it is a scheme-independent
physical quantity. (Unless indicated otherwise hereafter,
the scheme taken for the $b_n$ and resultant $\beta_{n\ell}$, 
$\alpha_{IR,n\ell}$, and $\gamma_{IR,n\ell}$ with $n \ge 3$ is the 
$\overline{\rm MS}$ scheme.)

Our previous work showed the usefulness of higher-loop
calculations of $\gamma_{IR,n\ell}$.  For example, for a (vectorial) SU(3)
gauge theory with $N_f=12$ massless Dirac fermions, the values of
$\gamma_{IR,n\ell}$ at the two-loop, three-loop, and four-loop level were found
to be 0.773, 0.312, and 0.253, respectively \cite{bvh,ps}. Our four-loop
result, $\gamma_{IR,4\ell}$, is in good agreement with the fully
nonperturbative lattice calculations $\gamma_{IR} = 0.27 \pm 0.03$
\cite{hasenfratz1}, $\gamma_{IR} \simeq 0.25$ \cite{hasenfratz2}, and
$\gamma_{IR}=0.235 \pm 0.046$ \cite{lombardo},\cite{lgtreviews}.  These
measurements are part of an intensive lattice program to elucidate the
properties of asymptotically free gauge theories with various fermion contents,
in particular, those exhibiting quasiconformal behavior; besides 
their intrinsic field-theoretic interest, such theories might play a role in 
physics beyond the Standard Model \cite{lgtreviews}. Similar 
agreement was found for four-loop calculations in other schemes
\cite{sch}-\cite{gracey2015}. An iterative method to calculate $\gamma_{IR}$ in
a scheme-independent manner has been presented in \cite{tr2016}. It allows for
a direct comparison of perturbative methods with exact results in
$\mathcal{N}=1$ supersymmetric QCD, for which it was shown that $\gamma_{IR}$
is very well described already at a few loops level throughout the entire
conformal interval.

In the UV to IR evolution, as $\mu$ decreases, $\alpha(\mu)$ approaches the IR
zero in $\beta$. If this zero occurs at relatively weak coupling, it can be an
exact IR fixed point (IRFP) of the RG, and the corresponding IR phase is a
chirally symmetric, deconfined non-Abelian Coulomb phase (NACP, conformal
interval). If the IR zero in $\beta$ occurs at a sufficiently large value of
$\alpha$, then the IR phase has confinement and spontaneous chiral symmetry
breaking (S$\chi$SB) associated with a nonzero bilinear fermion condensate
formed at a scale $\Lambda$. In this case, the fermions gain dynamical masses
and are integrated out of the low-energy effective theory applicable for 
$\mu < \Lambda$.  The IR zero in $\beta$ is then only an approximate IRFP and
similarly, $\gamma_{IR}$ is only an effective quantity describing the RG flow
near this approximate IRFP.

We next describe the behavior of $b_5$ as a function of $N_f$ (for the behavior
of $b_3$ and $b_4$, see \cite{bvh,bc}.)  As $N_f \in I_{AF}$ increases from 0,
$b_5$ initially decreases through positive values, reaches a minimum at
$N_f=6.074$ \cite{nfintegral}, where $\bar b_5=0.640 \times 10^{-3}$, and then
increases.  For all $N_f \in I_{AF}$, $b_3$ is negative-definite, while 
$b_4$ and $b_5$ are positive-definite.  We list values of the $\bar b_\ell$ for
$1 \le \ell \le 5$ in Table \ref{bnbar_values}.

\begin{table}
\caption{\footnotesize{Values of the $\bar b_\ell$ for
$1 \le \ell \le 5$ as a function of $N_f$, with
$b_\ell$ for $\ell=3,4,5$ calculated in the $\overline{\rm MS}$ scheme.}}
\begin{center}
\begin{tabular}{|c|c|c|c|c|c|} \hline\hline
$N_f$ & $\bar b_1$ & $\bar b_2$ & $\bar b_3$ & $\bar b_4$ & $\bar b_5$
\\ \hline
0   &  0.875  & 0.646    & 0.720     & 1.173 & 1.714   \\
1   &  0.822  & 0.566    & 0.582     & 0.910 & 1.175   \\
2   &  0.769  & 0.485    & 0.450     & 0.681 & 0.744   \\
3   &  0.716  & 0.405    & 0.324     & 0.485 & 0.416   \\
4   &  0.663  & 0.325    & 0.205     & 0.322 & 0.186   \\
5   &  0.610  & 0.245    & 0.091     & 0.194 & 0.0494  \\
6   &  0.557  & 0.165    & $-0.016$  & 0.099 & 0.000866 \\
7   &  0.504  & 0.084    & $-0.118$  & 0.039 & 0.0354  \\
8   &  0.451  & 0.004    & $-0.213$  & 0.015 & 0.1475  \\
9   &  0.398  & $-0.076$ & $-0.303$  & 0.025 & 0.332  \\
10  &  0.345  & $-0.156$ & $-0.386$  & 0.072 & 0.583  \\
11  &  0.292  & $-0.236$ & $-0.463$  & 0.154 & 0.894  \\
12  &  0.239  & $-0.317$ & $-0.534$  & 0.273 & 1.261  \\
13  &  0.186  & $-0.397$ & $-0.599$  & 0.429 & 1.676  \\
14  &  0.133  & $-0.477$ & $-0.658$  & 0.622 & 2.134  \\
15  &  0.080  & $-0.557$ & $-0.711$  & 0.852 & 2.628  \\
16  &  0.0265 & $-0.637$ & $-0.758$  & 1.121 & 3.152  \\
\hline\hline
\end{tabular}
\end{center}
\label{bnbar_values}
\end{table}

For our analysis of the IR zero of $\beta$, it is convenient to extract a
prefactor and define a reduced $n$-loop beta function as 
\beq
\beta_{r,n\ell} \equiv \frac{\beta_{n\ell}}{-2 \alpha^2 \bar b_1} = 
1 + \sum_{\ell=2}^n \bar \rho_\ell \, \alpha^{\ell-1}
\label{betar_nloop}
\eeq
where $\bar\rho_\ell=\bar b_\ell/{\bar b_1}$. The equation $\beta_{r,n\ell}=0$
determines the IR zero and is a polynomial equation of degree $n-1$ in 
$\alpha$. Among the $n-1$ roots,
the smallest positive (real) root, if there is such a root, is
$\alpha_{IR,n\ell}$.  The nature of the roots at the $n=3$ and $n=4$ loop level
has been discussed in \cite{bvh,ps}.  

We present our results for
$\alpha_{IR,5\ell}$ in Table \ref{alfir_nloop_values}.  We 
begin the discussion at the upper end of the interval $I_{IRZ}$. 
For $14 \le N_f \le 16$, we find that 
$\alpha_{IR,5\ell}$ is close to, and slightly larger than, 
$\alpha_{IR,4\ell}$. For $N_f=13$,
$\alpha_{IR,5\ell}$ is about 20 \% larger than $\alpha_{IR,4\ell}$.  If 
$9 \le N_f \le 12$, we find that the five-loop beta function (in the 
$\overline{\rm MS}$ scheme, with $b_5$ from \cite{bck}) has no physical IR 
zero; instead, the roots of the quartic polynomial $\beta_{r,5\ell}$ consist 
of two complex-conjugate (c.c.) pairs.  This is a surprising result, since at
all of the lower-loop orders, namely $n=2$, $n=3$, and $n=4$, for 
$N_f \in I_{IRZ}$, the $n$-loop beta functions (in this $\overline{\rm MS}$
scheme and also other schemes \cite{sch}-\cite{gracey2015}) 
have physical IR zeros
$\alpha_{IR,n\ell}$, and one would naturally expect that as one extends the
calculation of $\beta_{n\ell}$ to higher-loop order, this behavior would
continue.  Specifically, we find the following: 
$N_f=9  \Rightarrow \alpha_{IR,5\ell} = 0.863 \pm 0.515i$;
$N_f=10 \Rightarrow \alpha_{IR,5\ell} = 0.715 \pm 0.382i$; 
$N_f=11 \Rightarrow \alpha_{IR,5\ell} = 0.609 \pm 0.277i$; and
$N_f=12 \Rightarrow \alpha_{IR,5\ell} = 0.528 \pm 0.176i$. 
Although these roots are unphysical if $9 \le N_f \le 12$, the respective 
real parts are similar to lower-loop values; for example, 
${\rm Re}(\alpha_{IR,5\ell})=0.609$ for $N_f=11$, which is close to 
$\alpha_{IR,4\ell}=0.626$, etc. 
As $N_f$ increases in this interval $9 \le N_f \le 12$, the
real part and the magnitude of the imaginary part decrease, consistent with the
approach to the real value $\alpha_{IR,5\ell}=0.406$ at $N_f=13$.  Formally
extending $N_f$ to real numbers, we find that as
$N_f$ approaches the value $N_f \simeq 12.8944$ from below, the two
complex-conjugate roots approach the real axis, with the real part approaching
0.47, and for larger $N_f \in I_{IRZ}$, the two c.c. roots are replaced by two
real roots, which respectively decrease and increase from $\alpha_{IR,5\ell}
\simeq 0.47$ as $N_f$ increases beyond 12.8944.
At the next physical integer value, $N_f=13$, the lower root in this pair
occurs at $\alpha_{IR,5\ell}=0.406$, as listed in Table
\ref{alfir_nloop_values}, while the upper one occurs at 0.5195. 

\begin{table}
\caption{\footnotesize{Values of $\alpha_{IR,n\ell}$ as a function of 
$N_f$ for $N_f \in I_{IRZ}$ and loop order $2 \le n \le 5$. See text for 
discussion of $\alpha_{IR,5\ell}$ for $9 \le N_f \le 12$.}}
\begin{center}
\begin{tabular}{|c|c|c|c|c|} \hline\hline
$N_f$ & $\alpha_{IR,2\ell}$ & $\alpha_{IR,3\ell}$ & $\alpha_{IR,4\ell}$
& $\alpha_{IR,5\ell}$ \\ \hline
9   &  5.24   & 1.028  & 1.072  & $-$    \\
10  &  2.21   & 0.764  & 0.815  & $-$    \\
11  &  1.23   & 0.578  & 0.626  & $-$    \\
12  &  0.754  & 0.435  & 0.470  & $-$    \\
13  &  0.468  & 0.317  & 0.337  & 0.406  \\
14  &  0.278  & 0.215  & 0.224  & 0.233  \\
15  &  0.143  & 0.123  & 0.126  & 0.127  \\
16  &  0.0416 & 0.0397 & 0.0398 & 0.0398 \\
\hline\hline
\end{tabular}
\end{center}
\label{alfir_nloop_values}
\end{table}

A necessary condition for the perturbative calculation of the IR zero to be
reliable is that the magnitude of the fractional difference
\beq
\Delta_{IR;n-1,n} = 
\frac{\alpha_{IR,(n-1)\ell} - \alpha_{IR,n\ell}}{\frac{1}{2}[
\alpha_{IR,(n-1)\ell}+ \alpha_{IR,n\ell}]}
\label{fracdif}
\eeq
should be reasonably small and should tend to decrease with increasing loop
order, $n$ \cite{asymp}. 
We have calculated the various $\Delta_{IR,n-1,n}$ and list the
values in Table \ref{delta_nnprime}.  As is evident, this necessary condition
is satisfied if $14 \le N_f \le 16$. If $N_f=13$, then the requisite behavior
is observed for $\Delta_{IR;23}$ and $|\Delta_{IR;34}|$, but $|\Delta_{IR;45}|$
is actually about three times larger than $|\Delta_{IR;34}|$. For lower values
of $N_f \in I_{IRZ}$, the $|\Delta_{IR;n-1,n}|$ criterion is not applicable,
since $\beta_{5\ell}$ is complex.  

These results are a consequence of the properties of the relevant coefficients
$\bar b_n$ in $\beta$.  In general, if, as a function of $N_f \in I_{IRZ}$,
$|\bar b_n|$ becomes very small in magnitude, then the $n$-loop contribution to
$\beta$ will tend to be a commensurately small correction to the $(n-1)$-loop
beta function, so $\Delta_{IR;n-1,n}$ will also be small.  
As $N_f$ decreases from 16 to 9, 
$|\bar b_3|$ decreases by a factor of 2.5 and $\bar b_4$ decreases sharply, 
by a factor of 45.  This strong decrease in $\bar b_4$ means that although the
overall size of $\alpha_{IR,4\ell}$ increases as $N_f$ decreases in this
interval $I_{IRZ}$, the fractional difference $\Delta_{IR;3,4}$ remains small,
as is evident in Table \ref{delta_nnprime}.  In contrast, although $\bar b_5$
also decreases as $N_f$ decreases in $I_{IRZ}$, it is still considerably 
larger than $\bar b_4$, leading to the larger value of
$|\Delta_{IR;4,5}|$ observed for $N_f=13$. 

\begin{table}
\caption{\footnotesize{Values of $\Delta_{IR;n-1,n}$ as a function of 
$N_f$ for $N_f \in I_{IRZ}$. See text for discussion of 
$\Delta_{IR;4,5}$ for $9 \le N_f \le 12$.}}
\begin{center}
\begin{tabular}{|c|c|c|c|} \hline\hline
$N_f$ & $\Delta_{IR;2,3}$ & $\Delta_{IR;3,4}$ & $\Delta_{IR;4,5}$ \\ \hline
9   &  1.344  & $-0.04175$  & $-$    \\
10  &  0.971  & $-0.0642$   & $-$    \\
11  &  0.723  & $-0.0791$   & $-$    \\
12  &  0.537  & $-0.0785$   & $-$   \\
13  &  0.386  & $-0.0639$   & $-0.185$      \\
14  &  0.258  & $-0.0415$   & $-0.0404$    \\
15  &  0.146  & $-0.0185$   & $-0.00770$   \\
16  &  0.0461 & $-0.00255$  & $-0.000288$  \\
\hline\hline
\end{tabular}
\end{center}
\label{delta_nnprime}
\end{table}

Our calculation of $\alpha_{IR,5\ell}$ thus reveals new complexities with the
IR zero in $\beta$ for $N_f \in I_{IRZ}$ that were
not observed at lower-loop level and hence were not anticipated at five-loop
order, since one expects that (in a nonpathological scheme) calculations at
higher-loop order should exhibit greater stability than those at lower-loop
order \cite{asymp}.  In view of our finding, we
next make use of the powerful method of Pad\'e approximants (PAs) \cite{pades}
to study the IR zero in $\beta$ at the five-loop level.  The $[p,q]$ PA to
$\beta_{r,n\ell}$ is the rational function
\beq
[p,q]_{\beta_{r,n\ell}} =
\frac{1+\sum_{j=1}^p \, n_j \alpha^j}{1+\sum_{k=1}^q d_k \, \alpha^k}
\label{pqx}
\eeq
with $p+q=n-1$, where the $n_j$ and $d_j$ are $\alpha$-independent 
coefficients.  For a given
$\beta_{r,n\ell}$, there are thus $n$ PAs, namely the
set $\{ \ [n-k,k-1]_{\beta_{r,n\ell}} \ \}$ with $1 \le k \le n$. For $n=5$ 
loops, this is the set $\{ [4,0], [3,1], [2,2], [1,3], [0,4]\}$. The [4,0]
PA is just $\beta_{r,5\ell}$ itself, which we have already analyzed, and the
[0,4] PA has no zero and hence cannot be used for the analysis of the IR
zero of $\beta_{r,5\ell}$, which
leaves us with the remaining three PAs.  We have calculated and analyzed these.
If a $[p,q]_{\beta_{r,n\ell}}$ PA has a physical IR zero at this $n=5$ loop 
level, it is denoted as $\alpha_{IR,5\ell,[p,q]}$.  Clearly, if a PA has 
a pole closer to the origin (indicated as $pcl$) than
a zero, then this zero is not a reliable guide to the UV to IR evolution of the
theory from weak coupling.  Furthermore, a PA may contain an
essentially coincident pair of a zero and pole (indicated by $zp$); in this
case, the zero and pole factors cancel and may be neglected. 

We present the results of our Pad\'e analysis in Table \ref{alfir_5loop_pades}.
Importantly, we find that in several cases the PAs yield results for the IR
zero at the five-loop level that are physical and/or more stable than the zeros
of $\beta_{r,5\ell}$ themselves.  For $N_f=16$ and $N_f=15$, all of the three
$\alpha_{IR,5\ell,[p,q]}$ listed in Table \ref{alfir_5loop_pades} agree very
well with the respective values of $\alpha_{IR,5\ell}$, and this is also true
for $\alpha_{IR,5\ell,[2,2]}$ and $\alpha_{IR,5\ell,[1,3]}$ in the case of
$N_f=14$.  For $N_f=13$, the values of $\alpha_{IR,5\ell,[2,2]}$ and
$\alpha_{IR,5\ell,[1,3]}$ lie roughly midway between $\alpha_{IR,4\ell}$ and
$\alpha_{IR,5\ell}$.  For $9 \le N_f \le 12$, where there is no physical IR
zero of $\beta_{r,5\ell}$, at least one of the PAs, namely 
$[3,1]_{\beta_{r,5\ell}}$, yields
physical IR zeros, and the respective values of $\alpha_{IR,5\ell,[3,1]}$ are
reasonably close to, and somewhat smaller than, the corresponding values of
$\alpha_{IR,4\ell}$.  (PAs that yield negative or complex zeros are marked with
$-$.)  Thus, using the physical results from the Pad\'e approximants helps to
circumvent the problem with complex $\alpha_{IR,5\ell}$ in this lower region of
$I_{IRZ}$.

\begin{table}
\caption{\footnotesize{Values of $\alpha_{IR,n\ell,[p,q]}$ from $[p,q]$ 
Pad\'e approximants to $\beta_{r,5\ell}$, as a function of 
$N_f \in I_{IRZ}$, including comparison with $\alpha_{IR,4\ell}$ and 
$\alpha_{IR,5\ell}$. The
symbols (i) $zp$ and (ii) $pcl$ mean that the Pad\'e approximant has 
(i) a coincident zero-pole pair closer to the origin, (ii) a pole or
complex-conjugate pair of poles closer to the origin in the complex $\alpha$
plane. Entries with $-$ are unphysical.}}
\begin{center}
\begin{tabular}{|c|c|c|c|c|c|} \hline\hline
$N_f$& $\alpha_{IR,4\ell}$ & $\alpha_{IR,5\ell}$ & $\alpha_{IR,5\ell,[3,1]}$
& $\alpha_{IR,5\ell,[2,2]}$ & $\alpha_{IR,5\ell,[1,3]}$ \\ \hline
9   &  1.072  & $-$     & 1.02$_{zp}$     & $-$    & $-$       \\
10  &  0.815  & $-$     & 0.756$_{zp}$    & $-$    & $pcl$     \\
11  &  0.626  & $-$     & 0.563$_{zp}$    & $-$    & $pcl$     \\
12  &  0.470  & $-$     & 0.4075$_{zp}$   & 0.634  & 0.614     \\
13  &  0.337  & 0.406  & $-$              & 0.376  & 0.375     \\
14  &  0.224  & 0.233  & $-$              & 0.232  & 0.232     \\
15  &  0.126  & 0.12 7  & 0.127           & 0.127  & 0.127     \\
16  &  0.0398 & 0.0398 & 0.0398           & 0.0398 & 0.0398    \\
\hline\hline
\end{tabular}
\end{center}
\label{alfir_5loop_pades}
\end{table}

The anomalous dimension $\gamma_m$ has the series expansion 
$\gamma_m=\sum_{\ell=1}^\infty c_\ell a^\ell$. 
The $n$-loop $\gamma_m$ is $\gamma_{m,n\ell}=\sum_{\ell=1}^n c_\ell a^\ell$. 
The coefficient $c_1=8$ is
scheme-independent, while the $c_\ell$ with $\ell \ge 2$ are
scheme-dependent \cite{gross75}. In the $\overline{\rm MS}$ scheme, the 
$c_\ell$ have been
calculated up to $\ell=4$ \cite{c4} and recently to $\ell=5$ \cite{bckgamma};
e.g., $c_2=(404/3)- (40/9)N_f$, etc. 

As noted above, we define $\gamma_{IR,n\ell} = \gamma_{n\ell}$ evaluated
at $\alpha=\alpha_{IR,n\ell}$.  We calculate $\gamma_{IR,5\ell}$ here. For
$14 \le N_f \le 16$, we use our values of $\alpha_{IR,5\ell}$.  For $N_f=13$,
we use $\alpha=\alpha_{IR,5\ell,[1,3]}$ and for $10 \le N_f \le 12$ we use
$\alpha=\alpha_{IR,5\ell,[3,1]}$.  In both the chirally symmetric and chirally
broken IR phases, the IR value of $\gamma_m$ has the upper bound
\cite{gammabound} $\gamma_{IR,n\ell} < 2$. Since $\gamma_{IR,2\ell}$ violates
this for $N_f=10$ \cite{bvh}, we only show results for 
$11 \le N_f \le 16$. These are given in Table \ref{gamma_nloop_values}.  
For $N_f$ values where
the five-loop IR zero occurs at sufficiently weak coupling, our new five-loop
value for the anomalous dimension at this zero is close to the four-loop value.
In particular, our value $\gamma_{IR,5\ell}=0.255$ at $N_f=12$ is in good
agreement with lattice measurements of this quantity, as was our value
$\gamma_{IR,4\ell}=0.253$ in \cite{bvh}.

\begin{table}
\caption{\footnotesize{Values of the five-loop anomalous dimension for the
fermion bilinear, $\gamma_{IR,5\ell}$, evaluated at the IR zero of the
five-loop beta function, $\beta_{5\ell}$, as a function of $N_f$ for 
$11 \le N_f \le 16$, including comparison with lower-loop values of 
$\gamma_{IR,n\ell}$.}}
\begin{center}
\begin{tabular}{|c|c|c|c|c|} \hline\hline
$N_f$& $\gamma_{IR,2\ell}$ & $\gamma_{IR,3\ell}$ & $\gamma_{IR,4\ell}$
& $\gamma_{IR,5\ell}$ \\ \hline
11  & 1.61   & 0.439  & 0.250  & 0.294  \\
12  & 0.773  & 0.312  & 0.253  & 0.255  \\
13  & 0.404  & 0.220  & 0.210  & 0.239  \\
14  & 0.212  & 0.146  & 0.147  & 0.154   \\
15  & 0.0997 & 0.0826 & 0.0836 & 0.0843  \\
16  & 0.0272 & 0.0258 & 0.0259 & 0.0259  \\
\hline\hline
\end{tabular}
\end{center}
\label{gamma_nloop_values}
\end{table}

In summary, using the recently calculated five-loop term in the SU(3) beta
function from \cite{bck}, we have presented the first calculation of the
five-loop IR zero in the beta function for an SU(3) gauge theory and the first
five-loop calculation of the anomalous dimension of the fermion bilinear
operator at this IR zero.  

The research of T.A.R. and R.S. was supported in part by the Danish National
Research Foundation grant DNRF90 to CP$^3$-Origins at SDU and by the U.S. NSF
Grant NSF-PHY-13-16617, respectively.


\begin{thebibliography}{99}
\bibitem{fm}
%
It is straightforward to include fermion mass terms, but if a given fermion has
a mass $m$, it is integrated out of the effective field theory applicable at
scales $\mu < m$ and does not affect the further evolution to the IR.
\bibitem{nfintegral}
%
Here and elsewhere, when an expression is given for $N_f$
that formally evaluates to a non-integral real value, it is understood
implicitly that one infers an appropriate integral value from it.
\bibitem{irgen}
%
Refs. \cite{bvh}-\cite{bc}
studied the IR zero in $\beta$ for general $G$ and
fermion representation $R$. Here we restrict to 
$G={\rm SU}(3)$ and $R=$ fundamental.
\bibitem{asymp}
%
Since the perturbative series expansion for a function in quantum field theory
is not, in general, a Taylor series expansion with finite radius of 
convergence, but instead, only an asymptotic expansion, one cannot apply 
the Taylor-Maclaurin remainder theorem to bound the difference between the true
function and the $n$-loop series approximation. Here $\Delta_{IR,n-1,n}$ is 
only intended to give an estimate of the stability of the $n$-loop 
calculation.
\end{thebibliography}
\end{document}